# Helix: A Unified Programmable Accelerator for Diverse Krylov Subspace Solvers

Wei Chen
Yale University
wei.chen.wc659@yale.edu

Yizhi Yuan
Yale University
yizhi.yuan@yale.edu

Linghao Song
Yale University
linghao.song@yale.edu

***Abstract*—Domain-specific accelerators deliver high performance and energy efficiency through specialization but often lack support for extensible algorithm families. Existing FPGA accelerators for Krylov subspace methods, which iteratively solve sparse linear systems, typically support a single solver or require separate hardware configurations. Helix follows a systematic algorithm-family design methodology that first abstracts common computational primitives from diverse Krylov solvers and builds the architecture around efficient, programmable primitive kernels. Helix then introduces programmable datapaths to coordinate these primitives. Its architecture decouples hardware implementation from solver-specific execution, enabling HBM reuse and zero solver-switching latency. Finally, Helix provides a DSL compiler that efficiently maps different solvers onto the architecture while optimizing vector-operation performance for each solver. We prototype Helix on a Xilinx Alveo U55C FPGA and evaluate Helix's practicality and performance by using it to compile and execute eight solvers. Helix achieves up to 7.91× and 1.16× speedups over Xilinx XcgSolver and the state-of-the-art FPGA accelerator Callipepla, respectively, both of which support only JPCG. Across the eight solvers, Helix sustains 47.5%−96.9% of the throughput of an NVIDIA RTX PRO 6000 GPU while delivering 2.29 × −4.66× higher energy efficiency and 1.89 × −3.85× higher throughput per unit peak memory bandwidth. The programmable components incur 0.25%−0.57% execution-time overhead and 4.7% LUT and 4.0% FF overhead, demonstrating negligible overhead for programmability.

## I. INTRODUCTION

Domain-specific accelerators have transformed modern computing by delivering remarkable performance and energy efficiency through aggressive hardware specialization [1]–[4]. However, modern scientific and AI software is evolving toward increasingly rich algorithm ecosystems rather than isolated algorithms. Software readily accommodates this evolution through extensible libraries, allowing new algorithms and variants to be integrated rapidly and selected at runtime. Hardware, in contrast, evolves on a much slower timescale, where supporting a new algorithm often entails redesigning architectures, recompiling toolchains, and redeploying hardware. This growing mismatch raises a fundamental challenge for future accelerator design: accelerators should evolve from accelerating individual algorithms to efficiently supporting entire algorithm families.

Krylov subspace methods are a family of iterative algorithms for solving large sparse linear systems [5] which dominate in a wide range of applications like scientific AI and scientific computing [6]–[10]. However, efficiently executing these methods remains challenging for CPUs and GPUs. Although TPUs [11] and reconfigurable architectures [12] offer acceleration for sparse operations, translating kernel-level performance into efficient end-to-end solver execution remains difficult [13], [14]. This gap motivates solver-specific accelerators. Practical systems utilize different Krylov solvers for different matrix structures [15]–[17], motivating a programmable accelerator for the Krylov solver family.

Among existing platforms, FPGAs provide a highly customizable hardware substrate to be tailored to the characteristics of different algorithm families. Modern FPGA platforms further integrate high-bandwidth memory (HBM), making them particularly attractive for sparse linear algebra workloads whose performance is often limited by irregular memory accesses. There are also ASIC-based reconfigurable architectures for sparse applications [12], [18], [19]. However, ASIC designs are expensive, require lengthy fabrication, and often rely on simulation for near-term evaluation, whereas FPGAs are economical, offer shorter development cycles, and allow direct evaluation on real hardware.

Existing FPGA accelerators for Krylov methods continue to follow a solver-specific design philosophy, either specializing in one solver or switching between solver-specific implementations. For example, Xilinx's CGSolver [20] and the state-of-the-art JPCG accelerator Callipepla [13] support only the CG family, while Acamar [14] supports multiple Krylov solvers by instantiating solver-specific architectures and switching bitstreams between them. They tightly couple hardware architectures to individual solver implementations. As a result, supporting a new solver requires redesigning the hardware architecture or deploying a different hardware configuration. Furthermore, they allocate one or more physical HBM channels to one operand throughout execution, underutilizing memory bandwidth when some operands are inactive.

Therefore, it is challenging to provide both high performance and easy-to-use programmability for an algorithm family accelerator. To handle this, we propose a design methodology for algorithm family accelerators, which includes three key steps: **Step 1: abstract common primitives to decouple architecture from specific algorithms**, separate hardware specialization from algorithm-specific execution by abstracting common primitives, design efficient primitive kernels, and provide programmability at the level of primitives; **Step 2: coordinate primitives efficiently,** coordinate primitives to preserve performance for each solver through a customized and programmable datapath between different primitives; **Step 3: map algorithms to architecture efficiently**, map solvers to the architecture automatically with performance optimization techniques so that each solver fully and efficiently utilizes the architecture and users can use it conveniently.

Based on these steps, we present Helix, a unified programmable accelerator for diverse Krylov solvers. In step 1, we seek a common abstraction across diverse Krylov solvers and find they share a common set of primitives, including SpMV, vector operations, scalar operations, and control. Helix builds efficient and programmable kernels for these computational primitives. In step 2, Helix builds programmable datapaths between SpMV and vector-processing kernels, as well as among vector-processing kernels. In step 3, we develop a custom compilation pipeline tailored to Krylov solvers that searches for an optimal mapping based on each solver's data dependencies. Helix includes a unified architecture and efficient algorithm-to-architecture mapping compilation pipeline. The contributions of this work are as follows:

- We present Helix, a unified programmable accelerator architecture for diverse Krylov solvers. Helix abstracts Krylov methods into common primitives, implements efficient programmable kernels, and provides a programmable datapath for flexible reuse of intermediate data. Compared with previous FPGA accelerators, Helix enables HBM reuse across vectors from different stages, with zero solver-switching latency.
- We develop a Krylov-solver domain compiler that maps diverse solvers onto the unified architecture. The compiler translates Krylov solver descriptions written in Helix's C-like domain-specific language into executable instructions, and explores an optimized schedule with minimal stages, simplifying solver programming for users and providing high performance for each solver.
- In addition to HBM reuse, Helix also balances stage-based operation fusion with data parallelism, further improving memory bandwidth utilization. We search for an optimized Helix architecture based on quantitative evaluation: we combine an analytical performance model and the compiler to estimate the performance of each solver under different configurations.
- We conduct extensive evaluation on a Xilinx Alveo U55C FPGA and demonstrate that Helix achieves up to 7.91× and 1.16× speedups over Xilinx XcgSolver and the state-of-the-art FPGA accelerator Callipepla, respectively, both of which support JPCG only. We evaluate Helix's portability and performance by compiling and running eight different Krylov solvers. It sustains 47.5 −96.9% of the throughput of an NVIDIA RTX PRO 6000 GPU while delivering 2.29−4.66× higher energy efficiency and 1.89 −3.85× higher throughput per unit peak memory bandwidth. The programmable architecture incurs only 0.25 −0.57% execution-time overhead and 4.7% LUT and 4.0% FF overhead with negligible other resources.

## II. Background

### A. Krylov subspace methods

Krylov subspace methods are a family of iterative algorithms for solving large sparse linear systems. Although different solvers employ different recurrence relations for symmetric positive-definite, symmetric-indefinite, and nonsymmetric systems, they all iteratively construct the solution through sparse matrix-vector multiplications and vector updates [5].

As illustrated in Fig. 1, representative Krylov solvers, including Conjugate Gradient (CG) [21], Minimum Residual (MINRES) [22], and BiConjugate Gradient Stabilized (BiCGSTAB) [23], share the same set of computational primitives despite their different recurrence relations. These primitives include sparse matrix-vector multiplication (SpMV), vector operations, reduction operations (dot product and norm), and scalar control. This common computational structure naturally motivates a programmable accelerator based on configurable computational primitives. Helix adopts the Jacobi preconditioner, which computes $z = D^{-1}r$. Since the diagonal inverse can be precomputed, the operation is implemented as an element-wise vector multiplication rather than division.

**CG**

$$\textbf{while } \|\mathbf{r}\|_2 > \epsilon \textbf{ do}$$
$$\mathbf{q} \leftarrow A\mathbf{p}$$
$$\rho \leftarrow \mathbf{r}^T\mathbf{r}$$
$$\sigma \leftarrow \mathbf{p}^T\mathbf{q}$$
$$\alpha \leftarrow \rho/\sigma$$
$$\mathbf{x} \leftarrow \mathbf{x} + \alpha\mathbf{p}$$
$$\mathbf{r} \leftarrow \mathbf{r} - \alpha\mathbf{q}$$
$$\rho_{\text{new}} \leftarrow \mathbf{r}^T\mathbf{r}$$
$$\beta \leftarrow \rho_{\text{new}}/\rho$$
$$\mathbf{p} \leftarrow \mathbf{r} + \beta\mathbf{p}$$
$$\textbf{end while}$$

**BiCGSTAB**

$$\textbf{while } \|\mathbf{r}\|_2 > \epsilon \textbf{ do}$$
$$\rho \leftarrow \hat{\mathbf{r}}^T\mathbf{r}$$
$$\beta \leftarrow \frac{\rho}{\rho_{\text{old}}}\frac{\alpha}{\omega}$$
$$\mathbf{p} \leftarrow \mathbf{r} + \beta(\mathbf{p} - \omega\mathbf{v})$$
$$\mathbf{v} \leftarrow A\mathbf{p}$$
$$\sigma \leftarrow \hat{\mathbf{r}}^T\mathbf{v}$$
$$\alpha \leftarrow \rho/\sigma$$
$$\mathbf{s} \leftarrow \mathbf{r} - \alpha\mathbf{v}$$
$$\mathbf{t} \leftarrow A\mathbf{s}$$
$$\omega \leftarrow \frac{\mathbf{t}^T\mathbf{s}}{\mathbf{t}^T\mathbf{t}}$$
$$\mathbf{x} \leftarrow \mathbf{x} + \alpha\mathbf{p} + \omega\mathbf{s}$$
$$\mathbf{r} \leftarrow \mathbf{s} - \omega\mathbf{t}$$
$$\textbf{end while}$$

**MINRES**

$$\textbf{while } \eta > \epsilon \textbf{ do}$$
$$\mathbf{w} \leftarrow A\mathbf{v} - \beta_0\mathbf{v}_0$$
$$\alpha \leftarrow \mathbf{v}^T\mathbf{w}$$
$$\mathbf{w} \leftarrow \mathbf{w} - \alpha\mathbf{v}$$
$$\beta \leftarrow \|\mathbf{w}\|_2$$
Update rotations
$$\phi \leftarrow c\xi, \xi \leftarrow s\xi$$
$$\mathbf{p} \leftarrow (\mathbf{v} - \tau_2\mathbf{p}_1 - \tau_3\mathbf{p}_2)/\tau_1$$
$$\mathbf{x} \leftarrow \mathbf{x} + \phi\mathbf{p}$$
$$\eta \leftarrow |\xi|/\|\mathbf{b}\|_2$$
Update vectors
$$\beta_0 \leftarrow \beta$$
$$\mathbf{v} \leftarrow \mathbf{w}/\beta$$
$$\textbf{end while}$$

**Vector:** vector operations **SpMV:** Sparse matrix vector multiplication
**Scalar:** coefficient update **Control:** branch, jump

Fig. 1. Core iterations of three representative Krylov solvers, CG, BiCGSTAB and MINRES. Different colors represent common computational primitives.

### B. Related Work

**1) FPGA-Based CG accelerators:** Early FPGA-based CG accelerators primarily targeted small-scale problems, limiting their applicability to large-scale real-world applictaions [24]–[27]. With the emergence of HBM-equipped FPGAs, recent accelerators have become capable of supporting large-scale sparse linear systems encountered in real-world applications. [28] is a benchmark-specific accelerator for HPCG that maximizes the data parallelism of individual operations. However, it frequently transfers intermediate vectors between the host CPU and HBM, making host-device communication a bottleneck. XcgSolver [20], [29], provided by the Xilinx Vitis HPC Library, is the first practical FPGA-based CG accelerator for large-scale sparse linear systems, storing different operands in different HBM channels to enable streaming execution. Callipepla [13] further reduces off-chip memory accesses through vector streaming reuse and improves memory bandwidth utilization with FP32 matrix read for SpMV.

However, both XcgSolver and Callipepla statically allocate HBM channels to operands for a fixed CG computation flow, resulting in underutilized memory bandwidth when operands are not active at some stages. Helix instead introduces a compilation framework that dynamically maps multiple logical operands across different stages to a set of same physical HBM channels, enabling the joint optimization of operation fusion and data parallelism for efficient HBM bandwidth utilization.

**2) Reconfigurable FPGA-based solver accelerator:** Acamar [14] pre-generates multiple bitstreams for different iterative solvers and SpMV implementations and dynamically switches among them in µs according to the input matrix. It primarily targets SpMV parallelism, while dense vector operations are implemented as static HLS kernels. In contrast, Helix supports multiple Krylov solvers on a unified architecture through instruction-level programmability with zero switching latency while optimizing the execution of the entire solver.

**3) ASIC-Based Reconfigurable architecture for Sparse Applications:** Alrescha [19] and Quartz [18] propose ASIC-based reconfigurable architectures for sparse applications, emphasizing programmability across different operations. However, efficient solver execution requires coordinating dependencies and data movement across operations, introducing mapping overhead and demanding specialized kernel interconnections for them. Helix addresses this challenge through specialized kernels and datapaths, together with optimized compilation techniques that coordinate operators while preserving easy-to-use programmability at the operation level.

**4) Other related works:** FPGA accelerators for SpMV, the primary kernel in Krylov solvers, have evolved from DDR-efficient streaming architectures and sparse matrix layouts [30], [31] to HBM-based designs exploiting memory-level parallelism [32]–[34]. Serpens [34] uses dependency-aware nonzero reordering to minimize pipeline stalls and stores matrix pointers in an HBM channel. LevelST [35] and Telos [36] focus on accelerating sparse triangular solvers supporting factorization-based preconditioners such as ILU and IC. Misam [37] selects Sparse GEMM dataflows and FPGA configurations based on input sparsity. La Superba [38] studies sparse-accelerator optimizations through self-comparison, relevant to evaluation methodology. BLR-Krylov [39] uses low-rank decomposition to reduce matrix access and Karp et al. [40] exploit spectral-element structure to accelerate a matrix-free CG solver, avoiding matrix storage and access. Domain-specific accelerators have been studied across FPGA, ASIC, CGRA, analog computing, wafer-scale processors, and processing-in-memory (PIM) platforms for applications ranging from machine learning [41]–[45] to graph analytics and graph neural networks [46]–[49], and scientific computing [18], [50]–[64]. To improve flexibility and programmability, prior work has explored programmable overlays [65]–[71] and reconfigurable architectures including coarse-grained reconfigurable architectures (CGRAs) [72]–[79]. Recent FPGA design automation research has improved accelerator development through task-parallel programming models, automated floorplanning, and high-level synthesis scheduling techniques [80]–[84].

## III. Helix Hardware Architecture

We abstract Krylov subspace methods and find four common primitives: vector operations, SpMV, scalar operations, and controls. Therefore, Helix's architecture is built around programmable kernels for these primitives, with programmable datapaths to coordinate their execution. Fig. 2 shows the overview of Helix. Each computation unit is driven by an instruction and then begins to execute when corresponding data arrives. Table I summarizes the ISA of Helix.

### *A. Architecture Components*

In this subsection, we introduce each component of Helix's architecture. We first introduce the primitive processing units, including the Vector Processing and Routing Unit (VPRU), the SpMV Processing Unit (SpMV PU), and the Scalar Processing Unit (Scalar PU). Then we introduce corresponding Load & Store Units followed by the programmable datapath among different primitives. Finally, we describe the hierarchical Controller.

**1) Primitive Processing Units:** As shown in Fig. 2(a), Helix provides three instruction-configurable processing units: VPRU, SpMV PU, and Scalar PU. To match the 512-bit HBM interface, each VPRU processes eight FP64 vector elements per cycle. Each SpMV PU processes eight 64-bit non-zero matrix elements per cycle, each represented by a 32-bit value and 32-bit metadata, and receives FP64 vectors. All VPRUs and SpMV PUs operate at FP64. VPRU: Vector Processing and Routing Unit. As shown in Fig. 2(c), VPRU integrates vector computation and operand routing in a unified hardware module which enables both efficient vector execution and direct forwarding of intermediate vectors between VPRUs. Helix uses the VECALU_CTRL instruction to configure each VPRU, specifying the operation type, source and destination operands, vector length, scalar coefficients, and bypass configuration. For vector computation, VPRU supports both element-wise and reduction operations: $\{\alpha x + \beta y, \alpha x - \beta y, \alpha x \odot \beta y;\ x^{\mathrm{T}} y;\ \|x\|_2\}$. As illustrated in Fig. 2(b), multiple VPRUs are replicated horizontally to exploit stage-level operation fusion and replicated vertically to exploit data parallelism. For operand routing, each VPRU receives vector streams from the vector load/store units or an upstream VPRU. It selects the operands required for local computation while forwarding generated results and unconsumed input streams to downstream VPRUs or back to the vector load/store units for memory writeback. The bypass fields identify vector streams that should be forwarded to downstream VPRUs or vector access units.

SpMV PU: SpMV Processing Unit. Helix adopts the window-based SpMV processing scheme of Serpens [34] and uses SPMV to configure SpMV PUs . Following Serpens, Helix processes SpMV in sliding windows as shown in Fig. 2(d). Each window covers a contiguous range of matrix columns and the corresponding segment of the input vector x. This segment is first loaded into replicated on-chip buffers. The PEs then process

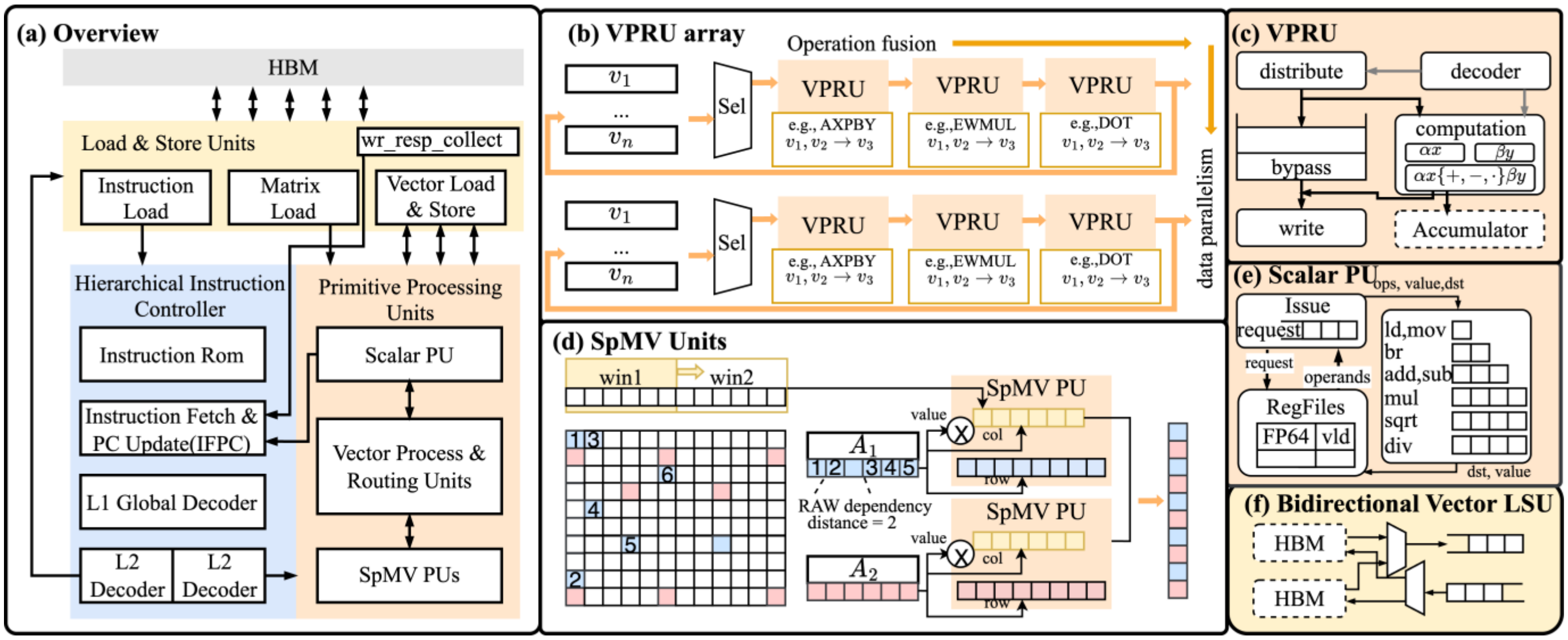

Fig. 2. Hardware architecture of Helix. (a) Overall architecture. (b) VPRU array, with stage-level vector operation fusion along the horizontal dimension and data parallelism along the vertical dimension. (c) Microarchitecture of a VPRU. (d) Sparse-matrix scheduling across two SpMV units. (e) Scalar Processing Unit. (f) Bidirectional vector LSU.

TABLE I
HELIX INSTRUCTION SET.

All instructions are 128 bits. The [127 : 123] 5-bit is opcode. The `term` bit is supplied by the PC at runtime. *Subopcodes.* `VECALU_CTRL.Op`: 1=ADD, 2=SUB, 3=MUL, 4=DOT, 5=NORM, 6=NORMDOT, 7=NOP. `SCALAR.code`: 0=MOV, 1=SLDI, 2=ADD, 3=SUB, 4=MUL, 5=DIV, 6=SQRT, 7=FABS, 8=OUT, 9=CMP_LE.

| **Opcode** | **Instruction** | **Fields** | **Function** |
|---|---|---|---|
| `00000` | `VEC_CTRL` | `vec_id[122 : 120], dir[119], rd[118], wr[117], region[116 : 109], base[63 : 32], len[31 : 0]` | Configures vector reads, writes, and routing. |
| `00001` | `MTX_CTRL` | `base[63 : 32], WIN_NUM[31 : 0]` | Configures matrix read streaming. |
| `00010` | `WR_RESP_ID` | `mask[115 : 112]` | Sets the write-response mask. |
| `00011` | `WR_RESP_RD` | – | Waits for vectors write completion. |
| `00100` | `VECALU_CTRL` | `blk_id[122 : 120], Op[119 : 116], vec1[115 : 113], vec2[112 : 110], dst[109 : 107], L2Dec_Reg_src1[106 : 105], L2Dec_Reg_src2[104 : 103], Bypass_Vecs_Mask[100 : 97], spmv_RdVec[96 : 94], spmv_WrVec[93 : 91], rd_src2[90], chunk_len[31 : 0]` | Configures vector operation, coefficient reg IDs, and data routing. |
| `00101` | `NORM_RD` | `L2Dec_Reg_dst[115 : 114]` | Write norm result into L2 Decoder local register. |
| `00110` | `DOT_RD` | `L2Dec_Reg_dst[115 : 114]` | Write dot result into L2 Decoder local register. |
| `00111` | `SPMV` | `WIN_NUM[63 : 32], chunk_len[31 : 0]` | Configures SpMV operation. |
| `01000` | `WR_RESIDUE` | `termination_flag[122], ScalarRF_src[115 : 114]` | Write residual and termination status to HBM. |
| `01100` | `SCALAR` | `code[119 : 116], dst[115 : 109], src1[108 : 102], src2[101 : 95], im_src[94 : 91], imm[63 : 0]` | Performs scalar arithmetic operations, including comparison. |
| `01101` | `SCALAR_RD` | `L2Dec_Reg_dst[115 : 114]` | Reads scalar into L2 Decoder local register. |
| `10000` | `BRANCH_LE_INT` | `offset[63 : 32], imm[31 : 0]` | Branches if $iter \leq imm$. |
| `10001` | `BRANCH_BNE_INT` | `offset[63 : 32], imm[31 : 0]` | Branches if $iter \neq imm$. |
| `10010` | `BRANCH_LE_FP` | `offset[63 : 32]` | Waits for scalar compare result as condition. |
| `10011` | `UPDATE_ITER` | – | Increments the iteration counter. |
| `10100` | `JAR_UNCOND` | `offset[63 : 32]` | Performs an unconditional relative jump. |
| `11000` | `RET` | – | Terminates execution. |

nonzeros within the current column range: for each $a_{ij}$, a PE uses column index j to fetch $x_j$ from the buffered segment and accumulates $a_{ij}x_j$ into the on-chip partial sum for row i.The window advances to the next column range while retaining these partial sums in a local accumulation buffer until all columns have been processed. Thus, irregular vector reads and accumulation accesses occur on chip, while off-chip data is streamed sequentially. Multiple SpMV PUs process different matrix rows parallelly while reusing the same input vector elements through a chain-based broadcast way.

Scalar PU: Scalar Processing Unit. As shown in Fig. 2(e), Helix employs an FP64 Scalar Processing Unit, consisting of an instruction issue module, a scalar register file with validity bits, and multiple arithmetic pipelines. The scalar instruction SCALAR specifies the operation type together with its source and destination registers. The issue module decodes instructions and checks the availability of operands, issuing an operation with operands to a pipeline when all source operands are valid. It also marks its destination register invalid to preserve data dependencies and dispatches it to the corresponding arithmetic pipeline. When execution completes, the Scalar PU writes the result back to the destination register and restores its validity. This validity-based dependency tracking allows the Scalar PU to issue scalar instructions immediately once their operands become ready while preserving execution order.

**2) Load and Store Units:** As shown in Fig. 2(a), Helix provides dedicated LSUs to manage data movement.

Vector Load and Store Unit. Helix uses VEC_CTRL instructions to configure the Vector LSUs. Helix defines two types of architectural vectors according to their access patterns within an execution stage: unidirectional vectors, which are accessed for either reading or writing, and bidirectional vectors, which support write-after-read or self-updating within one stage. One bidirectional architectural vector occupies twice as many HBM channels as a unidirectional architectural vector. But the latter trades access flexibility for the memory read and write bus, which is a routing bottleneck constrained by cross-logic-region wires of the FPGA. Accordingly, Helix provides two types of Vector LSUs. A unidirectional LSU connects to a single HBM channel and is configured for either reading or writing in each stage, whereas a bidirectional LSU connects to a pair of HBM channels as shown in Fig. 2(f). A logical unidirectional vector can be implemented by either a unidirectional or a bidirectional architectural vector, whereas a logical bidirectional vector is implemented by either one bidirectional architectural vector or a pair of unidirectional architectural vectors.

Matrix Load Unit. To support window-based SpMV processing, matrix nonzeros are preprocessed into per-window, per-PE streams. Within each stream, independent updates are interleaved to satisfy the read-after-write dependency distance for each buffer address, with no-ops inserted when necessary. For each window, the Matrix Load Unit sequentially reads these streams from HBM and delivers the scheduled nonzeros to the corresponding PEs. Helix configures the Matrix Load Unit through MTX_CTRL instructions, specifying the number of sliding windows and the base address of the first stream. Serpens [34] uses a separate HBM channel to store the start and end pointers of scheduled matrix streams for each window. Helix embeds these pointers as a head packet for each window, allowing each PE to obtain this metadata from its input stream.

**3) Hierarchical Instruction Controller.:** As shown in Fig 2(a), the Hierarchical Instruction Controller includes an Instruction ROM, an Instruction Fetch and Program Counter (IFPC) module, and two decoder levels. IFPC fetches instructions from the ROM and updates the PC according to sequential execution or branches. The L1 decoder predecodes the Opcode of instructions and dispatches them to two L2 decoders according to whether they are computation instructions or memory access instructions. The L2 decoders proceed independently, shortening the control critical path compared with a centralized decoder.

### B. *Coordination of Primitives*

**1) Data Reuse:** Helix provides data paths both between VPRUs and the SpMV Unit and between VPRUs, allowing vector data reuse. It uses VECALU_CTRL.spmv_WrVec to forward the results of VPRUs to SpMV units and vice versa, and uses VECALU CTRL.Bypass Vecs Mask to forward multiple vector streams to downstream VPRUs.

**2) Branch and synchronization for PC Updating. :** The PC has to stall for two types of instructions. One is BRANCH_LE_FP for FP64 comparison executed by the Scalar PU. Int comparison is finished inside the PC Update module. The other is WR_RESP_RD for waiting for completion of vector writes within a stage. Both these instructions are combined with another instruction. The compiler translates an FP64 conditional branch into two consecutive instructions: SCALAR with code=CMP_LE for the Scalar Unit and BRANCH_LE_FP for IFPC. Upon encountering BRANCH_LE_FP, IFPC stalls until it receives the result from the Scalar PU as shown in Fig. 2.(a). The compiler translates synchronization of vector write completion into two instructions: WR_RESP_ID for the write resp collect module, which collects write completion signals from all LSUs as shown in Fig. 2.(a), and WR_RESP_RD for IFPC. For example, a stage updates vectors 0 and 3, and the compiler emits a WR_RESP_ID instruction with bitmask 1001 for the write resp collect module after all VEC_CTRL instructions, and a WR_RESP_RD for IFPC after all data access and computing instructions within a stage to avoid stalling other instructions. After the write resp collect module receives completion signals from the corresponding LSUs, it will send a response to IFPC as shown in Fig. 2.(a).

## IV. HELIX'S COMPILATION PIPELINE

Helix employs a custom compiler implemented entirely in Python, without dependence on MLIR/LLVM infrastructure or reuse of existing transformation passes. The compiler maps the C-style Helix DSL into an execution binary for the architecture and includes an optimized stage schedule to minimize the total number of VecOps stages. Therefore, Helix provides not only programmability but also high performance.

**1) Overview of compilation pipeline:** As shown in Fig. 3(a), the compiler of Helix includes three phases. The first phase uses a lexer to identify tokens, a parser to organize them into operations, and semantic analysis to check their validity. Stage Scheduling then groups these operations into StageIR, an intermediate representation that maps high-level operations to

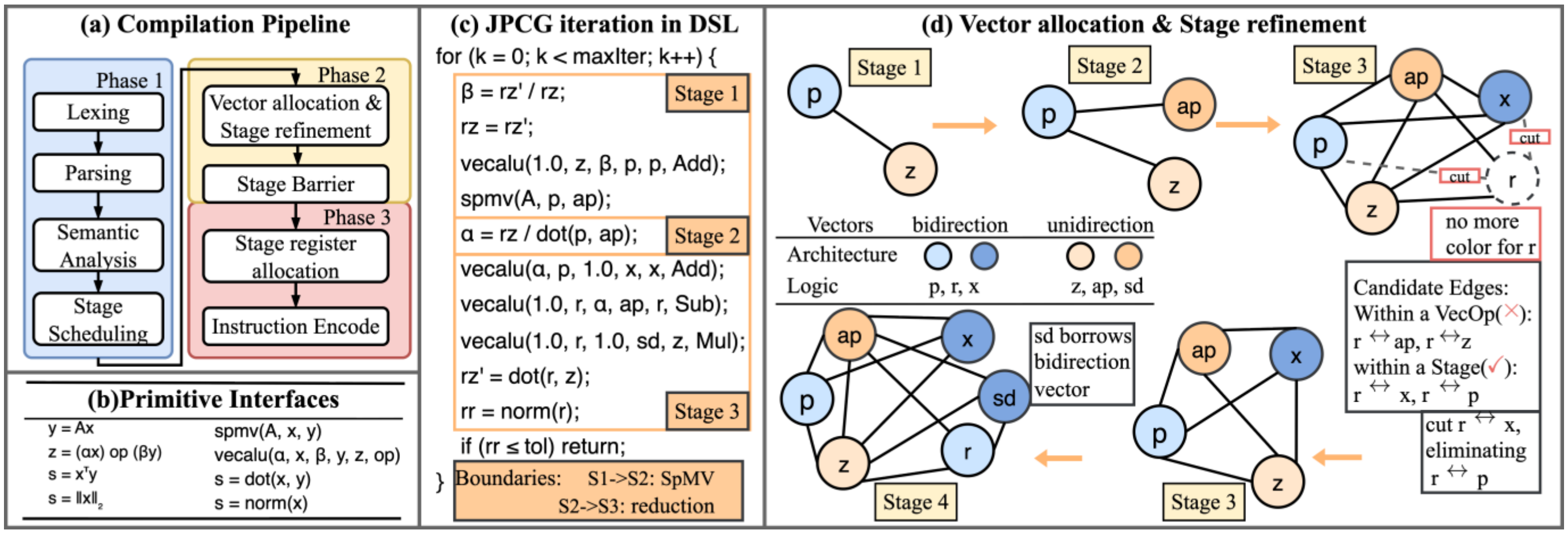

Fig. 3. Helix compilation pipeline. (a) Three compilation phases. (b) Primitive interfaces for SpMV and VecOps in the DSL. (c) JPCG iteration expressed in the C-style DSL and its stage schedule. (d) Vector allocation and stage refinement using graph coloring.

Helix's hardware execution stages. The second phase mainly allocates architecture vectors to logic vector operators for each stage. If it fails, it will recurrently use a shortest-path-based strategy to split stages further so that there is a feasible allocation for all logic vectors with the fewest stages. Then it adds stage barriers, like the WR_RESP_RD synchronization instructions. The third phase allocates registers for scalar operands, encodes the scheduled operations according to Helix's ISA, and emits the resulting binary instructions.

**2) DSL Practicality:** Helix's DSL preserves C syntax, expressing interfaces for primitive operations, SpMV, element-wise vecalu, dot, and norm as shown in Fig. 3.(b). It resembles calling a numerical library and provides practicality for users without knowledge of the architecture and compiler.

**3) Stage Scheduling:** Stage Scheduling presents operations in StageIR, mapping high-level operations to Helix's hardware execution stages. Organizing operations into stages exposes opportunities for stage-level operation fusion and intermediate vector reuse. Stage Scheduling separates execution stages by natural boundaries like reduction operations in Fig. 3(c).

**4) Graph-Coloring-Based Vector Allocation:** Given the execution stages, the compiler allocates logical vectors to architectural vectors, which are statically bound to physical HBM channels. To eliminate unnecessary inter-channel data movement, each logical vector retains the same architectural vector throughout the program. The allocation must ensure that (1) active logical vectors within a stage are assigned different architectural vectors, and (2) each logical vector is assigned consistently across all execution stages. The compiler formulates vector allocation as a graph-coloring problem, where logical vectors correspond to graph nodes and architectural vectors correspond to colors. Unlike conventional graph coloring, Helix supports heterogeneous architectural vectors. A logical bidirectional vector can be implemented by either one bidirectional architectural vector or two complementary unidirectional architectural vectors, whereas a logical unidirectional vector can use either type. Consequently, graph coloring is extended with heterogeneous resource constraints. As illustrated in Fig. 3(d), allocation succeeds for Stages 1 and 2 but fails for Stage 3. At Stage 3, no bidirectional architectural

**Algorithm 1** Shortest-Path Stage Schedule Refinement.

Initialize: the state set $\mathcal{S} \leftarrow \{P_0\}$ $P_0$ is initial schedule
**loop**
  Select $P \in \mathcal{S}$ with the minimum stages
  Remove $P$ from $\mathcal{S}$, construct graph for $P$
  **if** coloring succeeds **then**
    **return** $P$
  **end if**
  Generate all legal split candidates $\mathcal{C}$
  **for all** $c \in \mathcal{C}$ **do**
    $P' \leftarrow \text{ApplySplit}(P, c)$
    Insert $P'$ into $\mathcal{S}$
  **end for**
**end loop**

vector or feasible pair of unidirectional vectors is available for logical vector r. The compiler therefore invokes the stage refinement procedure described in the next subsection.

**5) Stage Refinement:** When graph coloring fails, no feasible vector allocation exists for the current stage schedule under the available architectural vector resources. Helix refines the schedule by splitting one execution stage into two while preserving all data dependencies, thereby eliminating only conflicts introduced by stage-level concurrency. Fig. 3(d) illustrates one such refinement. If the refined schedule remains infeasible, further stage refinement is performed until a feasible schedule is obtained.

To minimize the number of execution stages, Helix formulates stage refinement as a shortest-path search problem as shown in Algorithm 1, where each node represents a candidate schedule and each edge corresponds to a legal stage split. Since each split introduces exactly one additional execution stage, the shortest path directly yields the feasible schedule with the minimum number of execution stages.

To conclude, Helix's compiler provides programmability for users and optimizes the stage schedule for each solver to maintain high performance for the family accelerator.

## V. Architecture Optimizations

### A. Stage-Level Operation Fusion and Data Parallelism

Previous FPGA-based accelerators like those in [13], [14] bind one or more physical HBM channels to one logic vector during solver execution, underutilizing memory bandwidth when vectors are inactive in some stages. To handle this, Helix proposes to reuse HBM channels for logic vectors across different stages. However, different stages have imbalanced numbers of active vectors; for example, stage1 and stage2 have 15 and 4 logic vectors, respectively. Therefore, setting the number of HBM channels to 15 still leaves some HBM channels idle when executing stage2, while setting it to 4 causes stage1 to lose opportunities for stage-level operation fusion and data reuse. To solve this problem, we propose to incorporate data parallelism when deciding the HBM channels for each stage. Data reuse aims to reduce total data access, while data parallelism aims to improve memory bandwidth utilization. For example, we could use 3 sets of 5 HBM channels, providing both data reuse and parallelism in a balanced way. Nevertheless, the best design point needs quantitative analysis and varies with the data dependencies of different solvers. Therefore, we established a performance model which obtains the number of stages from the compiler to estimate the best performance of different solvers for each combination of data parallelism and HBM set size.

### B. Performance Model

We establish a performance model to estimate the best performance of different solvers across various workloads for each combination of data parallelism and HBM set size. Due to limited wires across the FPGA's logic regions, Helix uses two types of architectural vectors, unidirectional and bidirectional, to trade more HBM channels for feasible implementation on the FPGA with memory access direction flexibility. Therefore, the performance model uses the sizes of the unidirectional and bidirectional vector sets to represent the size of HBM sets.

For a matrix with M rows, K columns, and NNZ nonzeros, we use $N = \max(M, K)$ as the vector length. Let $V_{\mathrm{u}}$ and $V_{\mathrm{b}}$ count unidirectional and bidirectional architectural vectors, respectively, and $P_{\mathrm{vec}}$ count HBM channels and VPRUs per access direction. These vectors require $P_{\mathrm{vec}}$ and $2P_{\mathrm{vec}}$ channels, respectively, the latter supporting simultaneous reads and writes. Total vector-channel usage and vector-stage latency are $C_{\mathrm{vec}} = (V_{\mathrm{u}} + 2V_{\mathrm{b}})P_{\mathrm{vec}}, T_{\mathrm{vec}} = \left\lceil \frac{N}{8P_{\mathrm{vec}}} \right\rceil$ as each 512-bit HBM access transfers eight FP64 elements. Let $C_{\mathrm{mtx}}$ count HBM channels allocated to SpMV, each paired with a Matrix Load Unit and an SpMV PU, and $R_{\mathrm{tile}}$ count row tiles. Each tile reloads the full input vector, while all tiles collectively write the output vector once. Thus, tiled SpMV latency is $T_{\mathrm{spmv}} = R_{\mathrm{tile}}\lceil N/8 \rceil + \lceil \mathrm{NNZ}/(8C_{\mathrm{mtx}}) \rceil + T_{\mathrm{vec}}$, accounting for input-vector loading, matrix streaming, and output-vector processing, respectively.

Let $S_{\mathrm{vec}}$ count scheduled vector stages and $S_{\mathrm{spmv}}$ count SpMV operations per iteration. Each SpMV includes one input-vector stage and one output-vector stage, already accounted for in $T_{\mathrm{spmv}}$, leaving $S_{\mathrm{vec}}{}' = S_{\mathrm{vec}} - 2S_{\mathrm{spmv}}$ vector-only stages. Iteration latency is therefore $T_{\mathrm{iter}} = S_{\mathrm{vec}}{}' T_{\mathrm{vec}} + S_{\mathrm{spmv}} T_{\mathrm{spmv}}$, omitting negligible scalar and control latency.

TABLE II
ARCHITECTURAL CONFIGURATION FROM DESIGN SPACE EXPLORATION.

| Component | Configuration |
|---|---|
| Vector LSUs | bidir: 2 HBM CHs, unidir: 1 HBM CH<br>Helix: 2 × (2bidir + 2unidir) = 12 CHs<br>Helix-10CH: 2 × (1bidir + 3unidir) = 10 CHs |
| VPRU | 2 arrays × 4 VPRUs (FP64) |
| SpMV PUs | 16 (Rd FP32 Matrix, Rd FP64 vectors, MAC with FP64) |
| Matrix LU | 16 |
| Instr ROM | 8 KB = 128bit × 512 entries |
| Scalar RegFile | 128 entries (FP64) |

TABLE III
THE OPERATING FREQUENCY, UTILIZED MEMORY CAPACITY, PEAK MEMORY BANDWIDTH, AND RUNTIME POWER OF THE EVALUATED FPGA ACCELERATORS AND GPU.

| | Frequency | Memory | Bandwidth | Power |
|---|---|---|---|---|
| **Xcgsolver** | 300 MHz | 11.5 GB | 317 GB/s | 44 W |
| **Callipepla** | 225 MHz | 13 GB | 374 GB/s | 52 W |
| **Helix-10CH** | 226 MHz | 13 GB | 374 GB/s | 53 W |
| **Helix** | 226 MHz | 14 GB | 403 GB/s | 56 W |
| **RTX PRO 6000** | 1.85 GHz | 96 GB | 1.60 B/s | 270 W |

*Design Space Exploration with Fast Implementation Check*

With the performance model, Helix can find the best architectural configurations. However, FPGA implementation feasibility still needs checking with Vivado, which takes tens of hours. Helix therefore complements the analytical model with a fast implementation checker. The checker extracts the DSP, buffer, LUT and FF resources of each module and edge between modules from one placement and routing run. Then Helix quickly generates resource utilization for modules and edges across configurations and passes it to a pre-floorplan scheduler provided by AutoBridge[82], which can check feasibility in tens of minutes. AutoBridge can also check different designs, but limited parameterizability of the hardware code requires a complete implementation for each configuration. Moreover, generating utilization from code with AutoBridge adds tens of minutes of overhead. Therefore, we generate utilization directly from configurations with Python within a second. The selected architecture is evaluated for JPCG, MINRES, and BiCGSTAB and summarized in Table II.

## VI. Evaluation

### A. Evaluation Methodology

**1) Experimental Setup:** We evaluate Helix on Xilinx Alveo U55C FPGA. We compare against XcgSolver [20], the Xilinx HPC CG solver, and Callipepla [13], a state-of-the-art FPGA accelerator for JPCG, and compare an NVIDIA RTX PRO 6000 GPU for all eight solvers. To ensure a fair comparison, both baselines are implemented on same U55C FPGA. Based on the design-space exploration in Section V-B, Helix is configured with 16 SpMV processing elements, matching Callipepla, and two sets of vector processing units for data parallelism. We also implement Helix-10CH, with equal HBM channels and Callipepla to clarify our performance comes from our architecture and HBM reuse rather than additional HBM channels. Table II shows configurations Helix and Helix10CH. Table III summarizes the operating frequency, utilized memory capacity and bandwidth, and runtime power consumption of the evaluated FPGA accelerators and GPU. We use Vitis 2023.1 and TAPA [81] to implement Helix, Helix-10CH, and Callipepla. We implement GPU baseline on CUDA 12.8 using cuSPARSE for sparse matrix-vector multiplication and cuBLAS for vector operations .

TABLE IV
REPRESENTATIVE KRYLOV SOLVERS USED TO EVALUATE HELIX. ALL THESE SOLVERS ARE USED WITH JACOBI PRECONDITIONER.

| Solver | Recurrence Family | Matrix Class | Key Characteristics |
|---|---|---|---|
| **JPCG** [21] | Preconditioned CG | SPD | Short recurrence, reductions, and Jacobi scaling. |
| **PCR** [21] | Conjugate residual | SPD / symmetric | Residual-minimizing updates with additional live vector states. |
| **PCGS** [85] | Bi-Lanczos | Nonsymmetric | Shadow residuals and coupled vector updates. |
| **BiCGSTAB** [23] | Bi-Lanczos | Nonsymmetric | Coupled recurrences with shadow-residual updates. |
| **SYMMLQS** [22] | Symmetric Lanczos | Symmetric indefinite | Distinct scalar recurrences and vector dependencies. |
| **MINRES** [22] | Symmetric Lanczos | Symmetric indefinite | Residual-minimizing recurrence for indefinite systems. |
| **PQMRS** [86] | QMR-style | Nonsymmetric | Quasi-minimal residual smoothing and coupled recurrences. |
| **PIDR2** [87] | IDR(2) shadow space | Nonsymmetric | Multiple vector states, shadow-space projections, a small scalar solve. |

FPGA execution time is measured by timing the accelerator kernel from Xilinx runtime (XRT), whereas GPU execution time is measured using CUDA events cudaEventElapsedTime. For runtime breakdown, we use NVIDIA Nsight Systems to profile GPU execution, and we use different blocks of instructions to profile operations by XRT runtime. To reduce measurement noise, we repeat each GPU experiment ten times and report median value.

**2) Benchmarks:** To evaluate the programmability and generality of Helix, we implement eight representative Krylov solvers spanning six recurrence families, as summarized in Table IV. Together, they cover Symmetric Positive Definite (SPD), symmetric indefinite, and nonsymmetric linear systems while exhibiting diverse recurrence structures, and computational characteristics. Consequently, they require different combinations of vector, SpMV, and scalar operations, providing a comprehensive evaluation of Helix across a broad range of Krylov algorithms.

**3) Workloads:** Our evaluation focuses on the feasibility of deploying different Krylov algorithms on a shared architecture and their performance relative to corresponding GPU implementations. We therefore use a common test set of 29 sparse SPD matrices from the SuiteSparse Matrix Collection [88], covering structural mechanics, electromagnetics, geomechanics, biomechanics, fluid flow and heat transfer, and landscape ecology. The matrices span a wide range of problem sizes, with dimensions ranging from 2.9K to 1M and the number of nonzeros ranging from 99K to 40M. This diversity captures varying memory footprints and computational intensities, enabling a comprehensive evaluation across different workloads.

**4) Convergence and performance metrics:** Solving terminates when the absolute iterative residual norm falls below $10^{-12}$ or the iteration count reaches 20K. Performance is measured by solver kernel execution time or time per kernel. When both Helix and the baseline converge within 20K iterations, we compare their total solution time. However, if either does not converge, we compare kernel time per iteration to compare architectural performance. In practice, users can switch solvers without switching latency when the selected solver fails to converge, motivating Helix's support for multiple solvers without changing the bitstream.

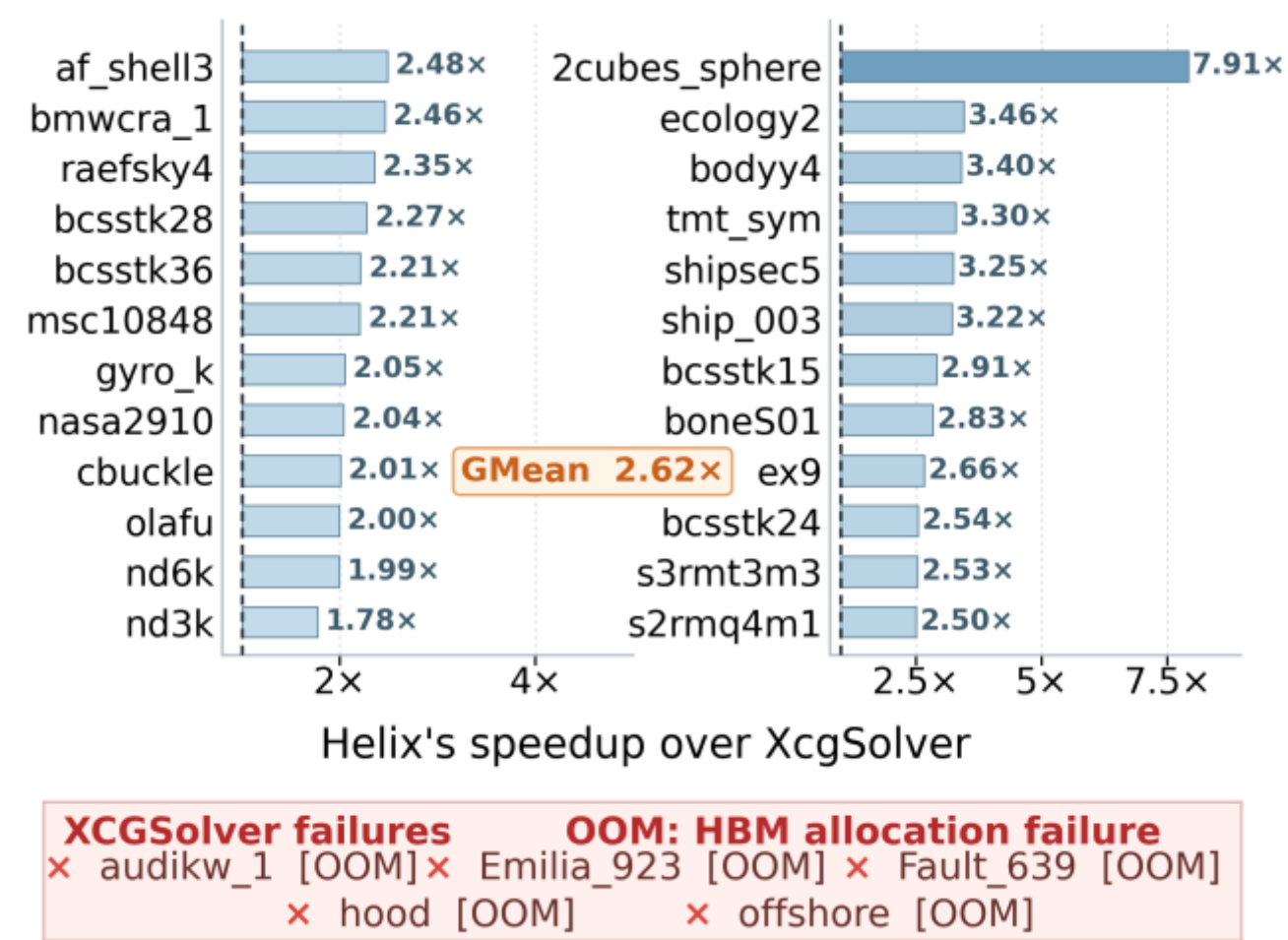

Fig. 4. Speedup of Helix over XcgSolver [20]. Helix achieves a maximum speedup of 7.91× and a geometric mean speedup of 2.62×. XcgSolver cannot complete five workloads because of insufficient memory capacity.

## B. *Performance Evaluation*

This subsection first compares Helix with XcgSolver and Callipepla, which support only JPCG. In contrast, Helix supports diverse Krylov solvers. We also use Helix-10CH which has same HBM channels of Callipepla to provide fair comparison. We then compare with GPU baseline.

**1) Compared with Xcgsolver:** As shown in Fig. 4, Helix performs better than XcgSolver on all workloads and achieves a maximum speedup of 7.91× and a geometric mean speedup of 2.62× over XcgSolver. XcgSolver is unable to complete five workloads because of insufficient memory capacity. XcgSolver's FP64 matrix data increase HBM capacity demand, while Helix uses FP32 matrix and stage-level fusion to reduce off-chip memory traffic.

**2) Compared with Callipepla:** As shown in Fig. 5(a), Helix achieves up to 1.52× and 1.16× speedup over Callipepla in VecOps execution and overall solver runtime, respectively. Helix performs better than Callipepla on 22 out of 29 workloads. For JPCG, the Helix compiler schedules the vector operations into four VecOps stages. Although this is one more stage than the three-stage execution in Callipepla, Helix employs two parallel VPRUs sets, providing higher vector-processing parallelism and yielding a theoretical upper-bound VecOps speedup of 1.50×. As shown in Fig. 5(a), the VecOps speedup of Helix generally increases with matrix size before stabilizing near the theoretical upper bound. For small matrices, the fixed HBM row-activation latency accounts for a relatively large fraction of the VecOps execution time, offsetting the latency reduction from parallelism. This is determined by HBM's physical characteristics rather than architectural design. Simply increasing parallelism for Callipepla's vectors also leads to this degradation. As the matrix size increases, the row-activation overhead is amortized over larger memory transfers, allowing the VecOps speedup to approach its theoretical limit. The slight overshoot observed for 2cubes sphere is likely caused by fixed runtime overheads, as the solver converges in only 33 iterations.

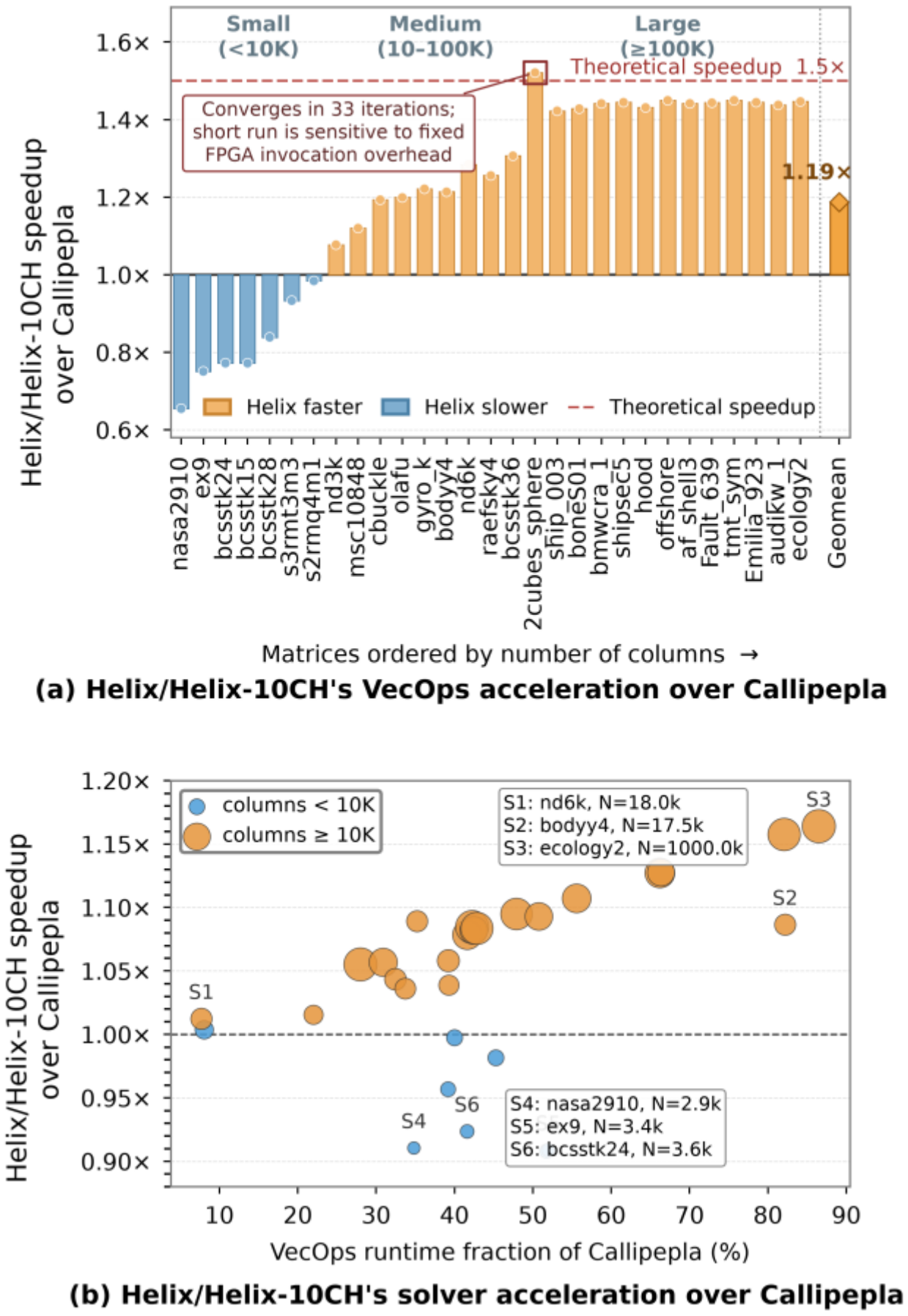

Fig. 5. Performance comparison of Helix/Helix-10CH with Callipepla [13]; Helix and Helix-10CH perform identically. (a) VecOps speedup versus matrix size; the dashed line marks the 1.50× theoretical bound. (b) Solver speedup versus Callipepla's VecOps runtime fraction.

Since both Helix and Callipepla employ the same number of parallel SpMV processing elements, their SpMV execution time is nearly identical. Consequently, the overall speedup of Helix comes from its HBM reuse and optimized stage schedules for VecOps. As shown in Fig. 5(b), workloads with a higher VecOps runtime fraction generally benefit more from Helix. For example, although nd6k and bodyy4 have comparable matrix dimensions, VecOps account for only 11.1% of the execution time in nd6k, compared with 83.4% in bodyy4. Accordingly, Helix achieves only a 1.2% speedup on nd6k, but 8.6% on bodyy4.

We compile JPCG for Helix-10CH and measure its on-board performance. The compiler produces the same number of stages for Helix-10CH as for Helix, and they achieve the same frequency. Therefore, Helix and Helix-10CH achieve identical performance for JPCG. This demonstrates that Helix's performance gain over Callipepla comes from Helix's architecture and compiler, which reuse HBM channels for vectors from different stages with an optimized stage schedule, improving memory bandwidth utilization for vector access compared with Callipepla. We select Helix rather than Helix-10CH to support JPCG, MINRES, and BiCGSTAB because Helix's additional bidirectional vectors benefit MINRES and BiCGSTAB, but do not benefit JPCG.

**3) Compared with GPU:** Fig. 6(a) shows Helix's speedup over an NVIDIA RTX PRO 6000 GPU across all matrices ordered by the number of matrix columns. Helix achieves up to 5.69× speedup on small workloads due to the GPU's kernel control overhead. As workload size increases, Helix's speedup decreases. This reflects the memory-bound nature of Krylov solvers. The GPU provides higher memory bandwidth than Helix (1.60 TB/s vs. 403 GB/s). Moreover, its larger on-chip memory allows accesses at 44.52 TB/s, whereas Helix must access corresponding data from HBM at 403 GB/s.

We profile execution time fractions of VecOps and SpMV of MINRES on the GPU and Helix as shown in Fig. 6(b) and Fig. 6(c). We find the overall performance improvement primarily comes from the substantially higher speedup of Helix for VecOps. We attribute this to Helix's optimized stage schedule. We further demonstrate this by comparing the reduced stages and corresponding speedup of each solver at Helix with GPU in the subsequent subsection.

**4) Impact of Stage-Based Operation Fusion:** Fig. 6(d) summarizes the speedup across eight solvers, with geometric means ranging from 0.475× to 0.969×. To explain this variation, we quantify Helix's operation fusion using the stage reduction ratio: the number of vector operations executed on the GPU divided by the number of VecOps stages generated by Helix. Fig. 6(e) summarizes this ratio across Krylov solvers. A higher ratio indicates more vector operations are fused into one stage, increasing operation-level parallelism and reducing memory accesses. Although the compiler optimizes stage-based operation fusion, its effectiveness depends on each solver's dependency graph. SYMMLQS benefits most, achieving a stage reduction ratio of 3.17× by reducing 19 vector operations per iteration to six stages. In contrast, PIDR2 benefits least, achieving a ratio of 1.30×. Its complex dependency graph limits fusion opportunities, reducing 39 vector operations to 30 stages. With limited VecOps fusion potential and three SpMV operations per iteration, PIDR2 benefits less from Helix, yielding the lowest solver speedup.

**5) Performance Summary:** Our performance analysis demonstrates that Helix's performance gain primarily comes from its vector-operation optimizations. Previous FPGA accelerators, such as Callipepla, maximize intermediate data reuse but use fixed mappings from one logical vector to one HBM channel throughout execution, leaving some channels idle when their associated vectors are not accessed. Helix addresses this limitation by dynamically mapping multiple logical vectors

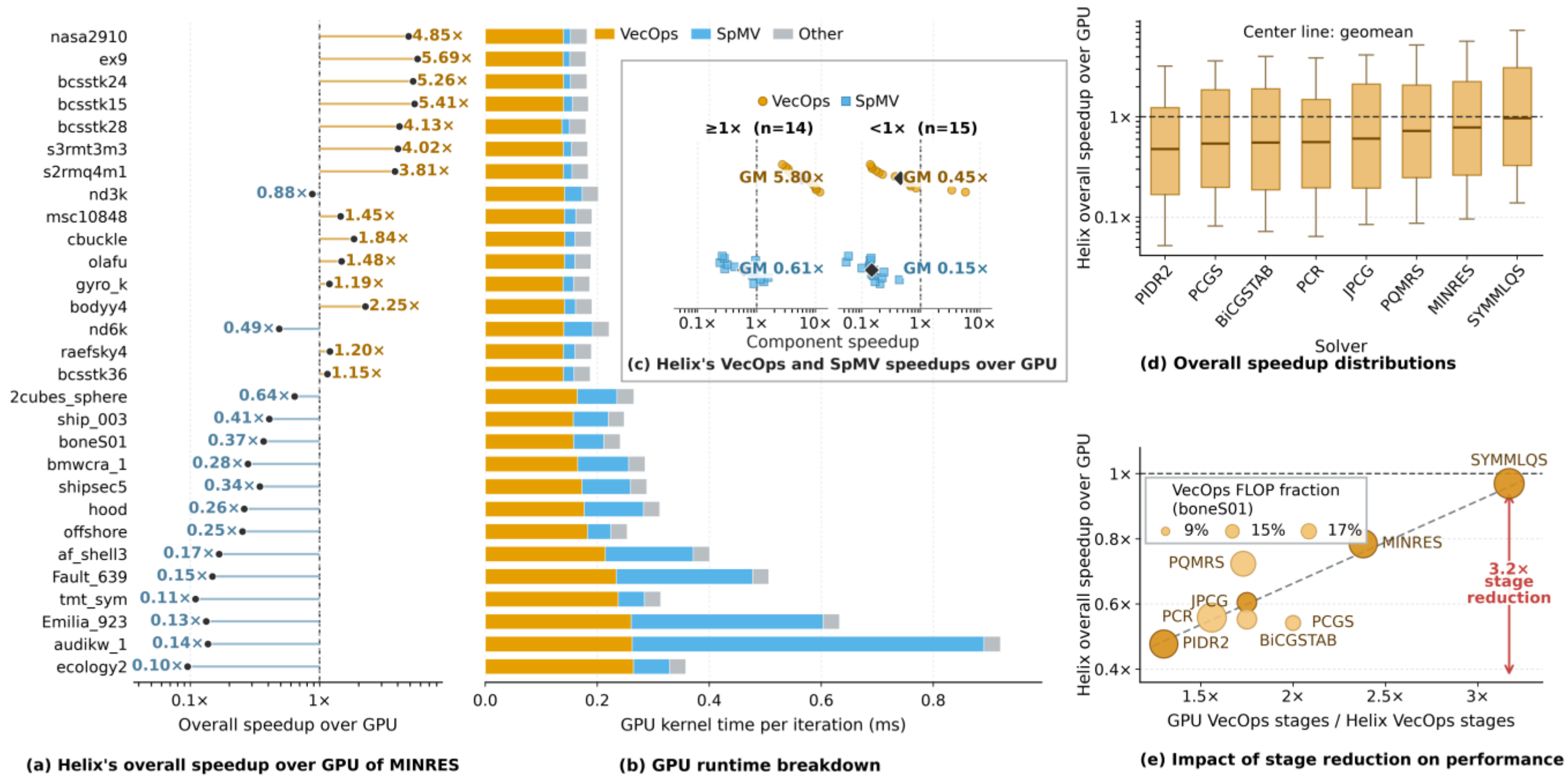

Fig. 6. Performance comparison between Helix and the GPU. (a) Speedup of Helix over the GPU across 29 matrices for MINRES. (b) GPU runtime breakdown for MINRES. (c) VecOps and SpMV speedups over the GPU. (d) Overall speedup distributions across eight Krylov solvers. (e) Stage reduction ratio and speedup, showing that solvers with more effective operation fusion generally achieve higher speedups.

TABLE V
KERNEL THROUGHPUT, ENERGY EFFICIENCY, AND THROUGHPUT NORMALIZED BY PEAK MEMORY BANDWIDTH.

| Solver / System | Throughput (GFLOP/s) | | | | | | | Energy Eff. (GFLOP/J) | | | Throughput/BW (FLOP/B) | | |
|---|---|---|---|---|---|---|---|---|---|---|---|---|---|
| | Reference | | | Helix | | | GM Ratio Helix/Ref | Reference | Helix | GM Ratio Helix/Ref | Reference | Helix | Ratio Helix/Ref |
| | Min | Max | GM | Min | Max | GM | | GM | GM | | GM | GM | |
| FPGA-XcgSolver | 2.46 | 23.16 | 9.21 | 12.66 | 43.33 | 24.66 | 2.62×* | 0.209 | 0.440 | 2.06×* | 0.0290 | 0.0612 | 2.06×* |
| FPGA-Callipepla | 11.37 | 42.82 | 23.72 | 12.66 | 43.33 | 24.66 | 1.04× | 0.456 | 0.440 | 0.96× | 0.0634 | 0.0612 | 0.97× |
| GPU-JPCG | 4.12 | 233.84 | 40.81 | 12.66 | 43.33 | 24.66 | 0.604× | 0.151 | 0.440 | 2.91× | 0.0256 | 0.0612 | 2.39× |
| GPU-PCR | 2.83 | 224.23 | 32.49 | 8.79 | 39.72 | 18.13 | 0.558× | 0.120 | 0.324 | 2.70× | 0.0203 | 0.0450 | 2.22× |
| GPU-PCGS | 4.72 | 234.30 | 44.53 | 11.56 | 43.20 | 24.07 | 0.540× | 0.165 | 0.430 | 2.61× | 0.0279 | 0.0597 | 2.14× |
| GPU-BiCGSTAB | 4.44 | 240.03 | 45.32 | 12.94 | 43.39 | 25.02 | 0.552× | 0.168 | 0.447 | 2.66× | 0.0284 | 0.0621 | 2.19× |
| GPU-SYMMLQS | 1.82 | 201.90 | 22.88 | 12.19 | 41.82 | 22.17 | 0.969× | 0.085 | 0.396 | 4.66× | 0.0143 | 0.0550 | 3.85× |
| GPU-MINRES | 2.00 | 205.00 | 24.05 | 9.21 | 40.22 | 18.85 | 0.783× | 0.089 | 0.337 | 3.79× | 0.0151 | 0.0468 | 3.09× |
| GPU-PQMRS | 2.19 | 210.64 | 26.29 | 9.09 | 40.53 | 19.00 | 0.723× | 0.097 | 0.339 | 3.49× | 0.0165 | 0.0471 | 2.85× |
| GPU-PIDR2 | 3.19 | 232.74 | 35.35 | 7.81 | 38.95 | 16.81 | 0.475× | 0.131 | 0.300 | 2.29× | 0.0221 | 0.0417 | 1.89× |
| Helix-10CH Vs Callipepla, Thourghput of Helix-10CH are same as Helix. | | | | | | | | 0.456 | 0.465 | 1.02× | 0.0634 | 0.0659 | 1.04× |

*Notes.* For algorithm rows, ratios report Helix relative to the GPU. XcgSolver and Callipepla are JPCG-only baselines, and their ratios report Helix relative to the corresponding system. We also compare Helix-10CH with Callipepla. Throughput reports minimum, maximum, and geometric mean (GM). *Ratios of Helix over Xcgsolver do not include 5 OOM workloads of Xcgsolver.

different stages to the same set of HBM channels. Furthermore, Helix's architecture balances stage-based operation fusion with data parallelism to address the imbalance in vector set across stages, and its compiler finds an optimized schedule to minimize stages under the configuration. Compared with our cuBLAS/cuSPARSE GPU baseline, which can reuse HBM channels but does not support optimized stage-based operation fusion for intermediate data reuse in VecOps, Helix achieves better whole-solver performance on some small- to medium-sized workloads through its VecOps improvements.

### C. Throughput, Energy and Memory-Bandwidth Efficiency

As shown in Table V, compared with XcgSolver, Helix achieves 2.62× higher geometric-mean throughput, 2.06× higher geometric-mean energy efficiency, and 2.06× higher geometric-mean throughput normalized by peakmemory bandwidth. These improvements come fromstage-based operation fusion and

TABLE VI
COMPARE REDUCED STAGES AND SPEEDUP OF SEARCH BASED STAGE-MINIMIZE SPLITTING STRATEGY AND SIMPLE SPLITTING STRATEGY CUTTING MOST RECENT CONFLICT EDGE.

| Algorithm | Vector stages Simple → Search | Stage reduction ratio | Geomean speedup |
|---|---|---|---|
| JPCG | 4 → 4 | 0.00% | 1.00× |
| MINRES | 8 → 8 | 0.00% | 1.00× |
| BiCGSTAB | 8 → 8 | 0.00% | 1.00× |
| SYMMLQS | 9 → 6 | 33.33% | 1.233× |
| PCR | 11 → 9 | 18.18% | 1.129× |
| PCGS | 9 → 8 | 11.11% | 1.047× |
| PQMRS | 13 → 13 | 0.00% | 1.00× |
| PIDR2 | 31 → 30 | 3.23% | 1.013× |

Helix's compressed FP32 matrix representation compared with Xcgsolver's FP64, which reduces off-chipmemory traffic and improves effective memory-bandwidth utilization. Compared with Callipepla, Helix improves throughput by 4% while reducing both energy efficiency and throughput normalized by peak memory bandwidth by 4% and 3%. The lower efficiency reflects the cost of programmability introduced by the programmable controller and the additional HBM channels allocated to accelerate VecOps. Although Helix-10CH is programmable, it still achieves higher energy efficiency and throughput than Callipepla with same HBM channels, which indicates the efficiency of HBM reuse and optimized stage schedule.

Across evaluated solvers, Helix sustains 47.5%–96.9% of the GPU's throughput while delivering 2.29×–4.66× higher energy efficiency and 1.89×–3.85× higher throughput normalized by peak memory bandwidth. Despite having only about one-quarter of the GPU's peak memory bandwidth, Helix achieves higher bandwidth efficiency through stage-based operation fusion and customized dataflow execution.

### D. Evaluation of stage-splitting strategy

Helix's compiler searches all splitting candidates to find a schedule with minimal stages when vector-allocation conflicts occur. We compare it with a simple stage-splitting strategy which cuts the most recent conflict edge. Table VI compares the stages and performance of both strategies.For SYMMLQS and PCR, we find most of conflicts come from one frequently used vector to a lot of other vectors which will conflict in further stage. Using simple strategy does not consider resolving further conflict, while the searched-based strategy reserves all candidates to resolve future conflicts as well as current conflict. Thus, the search-based splitting strategy reduces stages by 33.33% and 18.18%, providing 1.233× and 1.129× speedups for SYMMLQS and PCR. Conflicts in other solvers involve relatively fewer vectors, so the gain of search-based splitting strategy is minimal.

### E. Latency and Hardware Overhead of Instruction Control

To measure the cost of programmability, we evaluate the programmable controller in isolation by replacing all computation units with lightweight bypass modules while preserving the overall architecture. This isolates the controller from the computation pipelines. As shown in Fig. 7, the controller contributes only 0.25%–0.57% of the average iteration time across all evaluated Krylov solvers. Table VII shows that the overhead of Instruction Control primarily consumes 2.24% LUTs and 1.05% FFs of FPGA, amounting to 4.7% and 4.0% compared with Helix, while requiring 0.07% DSPs and no URAM. The BRAMs consumption is 4 and comes from instruction ROM, negligible compared with Helix's BRAM usage. Overall, Helix's introduced programmability incurs minimal runtime overhead, and its hardware overhead mainly comes from LUTs and FFs.

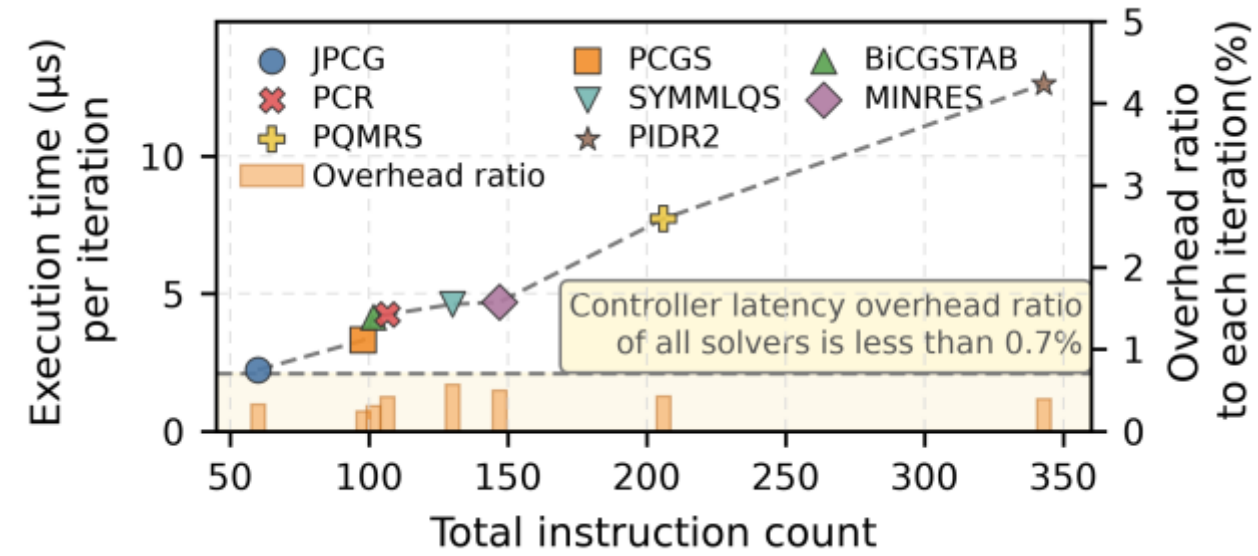

Fig. 7. Runtime overhead of the programmable controller across eight solvers. The measured controller latency scales with the total instruction count while accounting for only a small fraction of the average iteration time.

### F. Solver-switching Latency

When switching solvers for same workload, the initial residual and solution remain unchanged, thus the compiler can store them in same region and channels. Therefore, we only need to switch solver instructions on the FPGA. There are two ways to switch instructions, and the measured latency of either is zero. (1) The compiler compiles solvers together with a larger instruction ROM, and PC jumps directly to another instruction address. (2) The host CPU launches separate solver execution kernels, which execute in a pipelined manner. The host-to-FPGA transfer of the next solver's instructions overlaps with the FPGA-to-host transfer of the current solver's results. The absolute instruction transfer estimation is 0.458µs for PIDR2, which has the most instructions at 7.23KB, using the U55C's PCIe with 15.8 GB/s per direction bandwidth. It is fully hidden by the result transfer, even for the smallest workload's 23.2KB final solution, or for 1,000 FP64 residual values.

### G. Resource Utilization and Physical Layout

Table VII reports the post-route resource utilization of Helix and the FPGA baselines. Compared with the highly specialized Callipepla, Helix requires additional LUTs, FFs, and DSPs to support programmability. Fig. 8 shows the post-route layout of Helix. To exploit the available HBM bandwidth, SpMV processing elements, vector processing units, and HBM interfaces are distributed across multiple Super Logic Regions (SLRs), resulting in relatively high inter-SLR routing utilization, particularly between SLR0 and SLR1 (88.1%). To mitigate routing overhead, pairs of physical HBM channels share a read/write interface, reducing inter-SLR connections. Despite the aggressive bandwidth utilization, the design is successfully routed and meets timing constraints.

TABLE VII
POST-ROUTE RESOURCE UTILIZATION ON THE XILINX U55C FPGA.

| Design | LUT | FF | DSP | URAM | BRAM |
|---|---|---|---|---|---|
| XcgSolver | 487K (37.38%) | 783K (30.02%) | 1.2K (13.25%) | 128 (13.33%) | 590 (29.27%) |
| Callipepla | 511K (39.19%) | 554K (21.24%) | 1.9K (21.50%) | 384 (40.00%) | 712 (35.32%) |
| **Helix** | **626K (48.00%)** | **682K (26.17%)** | **3.1K (34.23%)** | **320 (33.33%)** | **716 (35.52%)** |
| *Helix's Ctrl* | 29K (2.24%) | 27K (1.05%) | 6 (0.07%) | 0 (0.00%) | 4 (0.20%) |
| *Helix-10CH* | 614K (47.12%) | 675K (25.90%) | 3.1K (34.23%) | 320 (33.33%) | 716 (35.52%) |

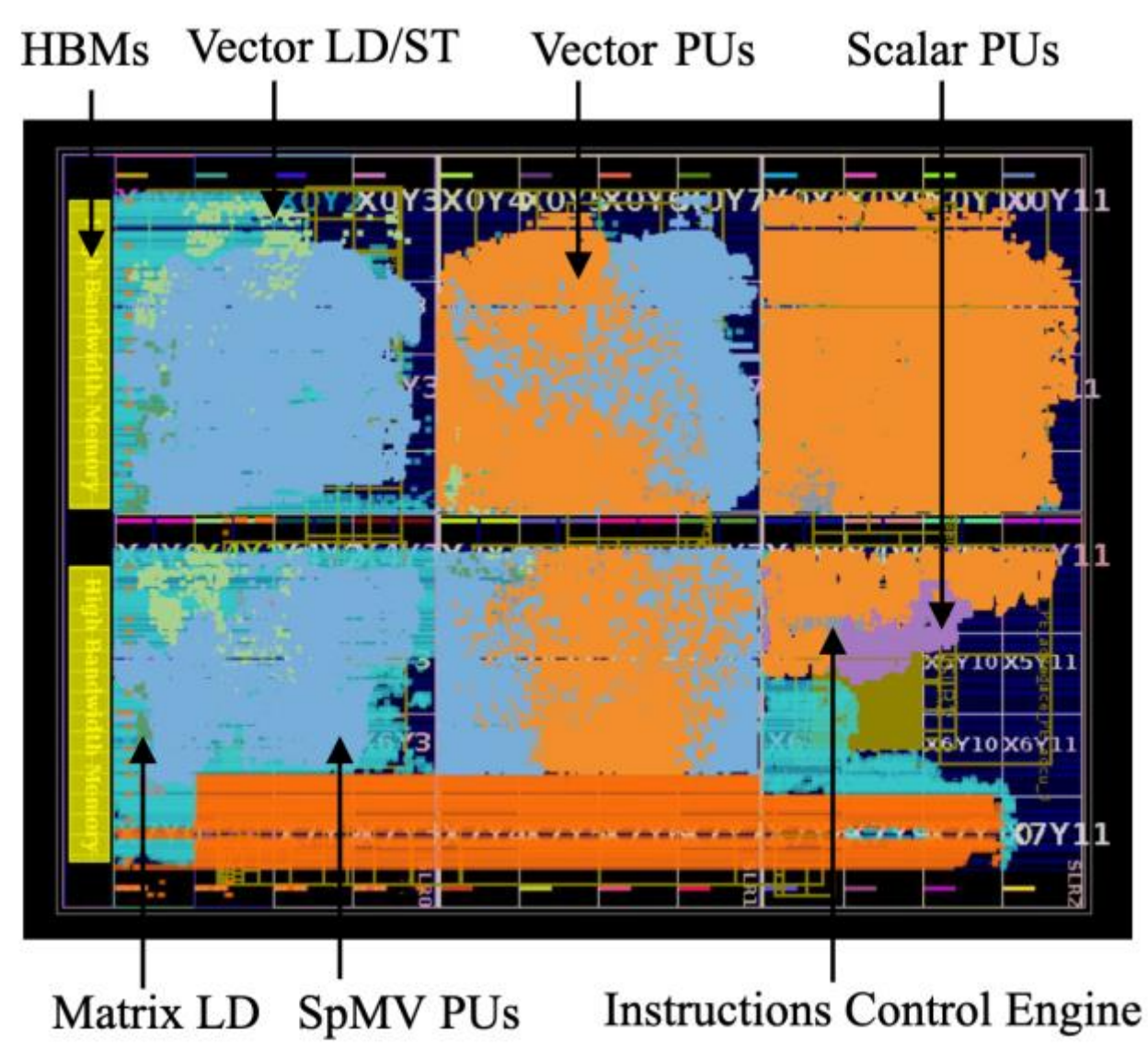

Fig. 8. Post-route layout of Helix on the Xilinx Alveo U55C FPGA.

### H. Convergence Behavior

Helix, Callipepla, Xcgsolver, and the GPU all use FP64 for vectors, scalars, and all arithmetic operations in VPRUs, the scalar PU, and the SpMV units. The only difference is that Helix and Callipepla store matrix values in FP32, whereas Xcg-solver and the GPU store them in FP64. Table VIII shows the convergence behavior of Helix and all baselines. The convergence behaviors of Helix, Callipepla and Xcgsolver are consistent: they converge or fail to converge on the same workload. This is because using FP64 for both arithmetic operations and the representation of vectors and scalars helps maintain numerical accuracy in residual computation and solution updates. Helix and the GPU present consistent convergence behavior on most matrices. However, for some matrices, like s3rmt3m3, the GPU converges while Helix does not, and vice versa. These differences can arise from different accumulation orders in Helix and the GPU, which propagate and accumulate floating-point rounding differences across iterations, affecting convergence. Helix's use of FP32 matrix values may also contribute. However, Helix's convergence behavior is similar to XcgSolver, which uses FP64 matrix values.

## VII. DISCUSSION

Comparison with CGRA and TPU We compare Helix with Plasticine [12], a representative CGRA and TPU [11] with SparseCore from two aspects: the implementation of Krylov

TABLE VIII
CONVERGENCE BEHAVIOR OF HELIX AND BASELINES.

| Baseline | Consistent behavior | | Inconsistent behavior | |
|---|---|---|---|---|
| | Both converge. | Neither converge. | Only Helix converge. | Only baseline converge. |
| XcgSolver | 19 | 5 | — | — |
| Callipepla | 24 | 5 | — | — |
| GPU-JPCG | 24 | 4 | — | s3rmt3m3 |
| GPU-MINRES | 23 | 4 | s3rmt3m3 | nd6k |
| GPU-BiCGSTAB | 11 | 15 | tmt_sym | nasa2910, shipsec5 |
| GPU-SYMMLQS | 24 | 5 | — | — |
| GPU-PCR | 24 | 4 | — | s3rmt3m3 |
| GPU-PCGS | 8 | 21 | — | — |
| GPU-PQMRS | 24 | 4 | — | s3rmt3m3 |
| GPU-PIDR2 | 16 | 13 | — | — |

Five XcgSolver workloads are excluded due to HBM allocation failures.

SpMV primitive and the specialized data path to coordinate operations. For SpMV, Plasticine and TPU's SparseCore both provide on-chip buffers and could potentially adopt Helix's scheduling strategy. However, data delivery may still be constrained by shared memory interfaces or configurable interconnects. Helix explicitly provisions dedicated paths between HBM and SpMV units, matching data delivery bandwidth to processing parallelism. For operations coordination, Helix directly supports SpMV–vector data reuse and fusion among vector operations with dedicated streaming connections. TPU's data exchange between SparseCore and VPU can involve HBM traffic, while its shared memory interfaces can limit operand delivery efficiency. Plasticine can realize similar data reuse with complex mapping, but configurable data paths may introduce multiple communication hops.

To conclude, a CGRA or TPU could integrate our specialized primitive kernels and scheduling strategy, but may coordinate operations less efficiently than Helix. We compare Helix's per-iteration execution time for JPCG with that of Google TPU v6e-1, using FP32 for SpMV and VecOps on the TPU. We use TPU-provided templates to optimize execution on its SparseCore and VPU. Helix achieves per-iteration speedups ranging from 1.36× (bodyy4) to 15.36× (offshore), with a geometric mean of 3.15×. We believe Plasticine could achieve performance comparable to Helix with the same HBM configuration, although overhead from datapath hops may make it slightly slower.

## VIII. CONCLUSION

This work demonstrates that hardware specialization can be extended from individual algorithms to algorithm families without sacrificing accelerator-level efficiency. Helix combines a unified programmable FPGA architecture and a Krylov solver domain compiler framework to support diverse Krylov solvers efficiently. Evaluation on a Xilinx Alveo U55C FPGA shows that Helix outperforms specialized FPGA baselines while sustaining 47.5% −96.9% of GPU throughput with up to 4.66× and 3.85× higher energy efficiency and throughput per unit peak memory bandwidth. Its programmable controller adds only 0.25%-0.57% runtime overhead. Helix provides both programmability and high performance, providing a methodology to design algorithm family accelerators.